# Scholars mobility and its impact on the knowledge producers' workforce of European regions

Márcia R. Ferreira[1], Juan Pablo Bascur[2] and Rodrigo Costas[3]

[1]*m.r.ferreira.goncalves@cwts.leidenuniv.nl;*
Centre for Science and Technology Studies (CWTS), Wassenaarseweg 62A, Leiden, 2333AL (The Netherlands)

[2]*j.p.bascur.cifuentes@cwts.leidenuniv.nl*
Centre for Science and Technology Studies (CWTS), Wassenaarseweg 62A, Leiden, 2333AL (The Netherlands)

[3] *rcostas@cwts.leidenuniv.nl*
Centre for Science and Technology Studies (CWTS), Wassenaarseweg 62A, Leiden, 2333AL (The Netherlands)
DST-NRF Centre of Excellence in Scientometrics and Science, Technology and Innovation Policy, Stellenbosch University (South Africa)

**Abstract**
Knowledge production increasingly relies on mobility. However, its role as a mechanism for knowledge recombination and dissemination remains largely unknown. Based on 1,244,080 Web of Science publications from 1,435,729 authors that we used to construct a panel dataset, we study the impact of inter-regional publishing and scientists' mobility in fostering the workforce composition of European countries during 2008-2017. Specifically, we collect information on scientists who have published in one region and then published elsewhere, and explore some determinants of regional and international mobility. Preliminary findings suggest that while talent pools of researchers are increasingly international, their movements seem to be steered by geographical structures. Future research will investigate the impact of mobility on the regional structure of scientific fields by accounting for the appearance and disappearance of research topics over time.

**Background**

After Europe's enlargement in 2004, a growing number of scholars had the opportunity to collaborate and live unconstrained by national borders. Some estimates indicate for instance that in 2010 there were about 1.59 million full-time researchers in the EU27 and that the number of researchers in the active population is increasing (See, for instance, More2 final report "Support for continued data collection and analysis concerning mobility patterns and career paths of researchers"). As national science systems become more globalized, they also become more dependent on the movement of researchers. This movement has been thought to be a driver of innovation and scientific breakthroughs, especially in receiving locations (Ganguli, 2015; Stephan & Levin, 2001).

However, one aspect of scientific mobility has been relatively neglected, namely, the mobility patterns of researchers across regional borders in Europe. As a result, the inter-regional structure of scientific mobility flows is still poorly understood. Yet, scholarly knowledge exchanges based on workforce mobility are vital to the transfer of tacit knowledge that cannot be transferred through formal communication channels (e.g., research articles, journals, books, e-mails) (Gertler, 2003). When mobile scholars move, they bring information and know-how, skills, and ideas that differ from natives and that are essential for knowledge recombination, interactive learning, and novelty. Based on previous theoretical and empirical work (e.g., Agrawal et al., 2006; Jaffe et al., 1993) we hypothesize that mobility is an effective mechanism for disseminating knowledge and capabilities across research institutions, fields and locations and is therefore important to our understanding of how fields evolve over time. Previous studies already point in that direction. Moser et al. (2014) showed that German-Jewish scientists fleeing from Nazi Germany into the US played a central role in the emergence of new chemistry sub-fields in which incumbents also participated. Ganguli (2015) showed that, after the 1991 breakdown of the Soviet Union, Russian scientists who migrated to the US were much more cited by US scientists than those who did not migrate. In a large-scale study of the Web of Science database, Sugimoto et al. (2017) found that scholars who were mobile (i.e., more than one affiliation to different countries) have approximately 40% more citations than non-mobile scholars.

Three exploratory questions will be asked at this preliminary stage: (1) How is scientific mobility distributed across European regions? (2) To what extent are international and regional mobility related? (3) Which regional mobility network patterns can be seen across countries? To answer these questions, we use Web of Science publications to construct a longitudinal data set for 264 regions from 32 European countries. In this research-in-progress paper, we briefly discuss the methodology, describe the data and discuss the preliminary findings and our plans for future research.

**The data**

Both publication data and regional data are used to compute regional mobility. Specifically, we reconstruct career trajectories of 1,435,729 distinct authors based on 1,244,080 publications (articles and reviews) published between 2008 and 2017. These data were collected from the CWTS in-house version of the Web of Science provided by Clarivate Analytics. Authors' profiles were isolated using the author-disambiguation algorithm developed by Caron and van Eck (2014). Since there is no consistent method for tracking scientific mobility, and categories of highly skilled workers are often too ambiguous to identify specific groups of scientists, we use author affiliations available in scientific publications. This is used to determine who moved between European regions, where and in what year. For each author we assign their affiliation addresses to the NUTS ('Nomenclature of Territorial Units for Statistics') classification structure, provided by Eurostat[1]. European countries differ in terms of size, population, number of researchers, availability of funding, and number of universities, therefore, we carry out the analysis at the regional level. We focus on regions with an average of at least 50 publications per year resulting in 264 NUTS2 European regions from 32 countries – EU-28 plus Norway, Iceland, Switzerland, and Turkey. Following on Robinson-Garcia et al. (2019) we define mobile scientists as those who have affiliations in at least two different regions either simultaneously (dual affiliation) or over the period of analysis (i.e., published in more than one region between 2008 and 2017). Accordingly, the dataset consists of 1,435,729 authors out of which about 14% (203,925) have published in more than one region. The mobility of scholars across regions is quantified by observed changes (movements) in the reported affiliation and corresponding region as stated by the scholars themselves in publication documents.

**Preliminary evidence**

*How is scientific mobility distributed across European regions?*

We here present preliminary evidence on regional mobility. A total 44% of scholars' movements involves NUTS2 regions belonging to the same country (regional mobility), while 55% of remaining movements involves international mobility. With respect to the distribution of movements of scholars across the European geography, 61 regions – out of 264 – concentrate 67% of scientists' movements. The countries with the highest numbers of mobile researchers are depicted in Figure 1. It follows that mobility flows of scholars tend to be clustered regionally within Europe. The regions in which we find heavy movement, such as Île-de-France, were also those where mobile researchers tended to locate more generally. Mobile researchers are noticeably absent from Eastern European regions, as these locations were also less likely to have lower scientific activity and lower levels of research funding.

---

[1] https://ec.europa.eu/eurostat/web/nuts/background

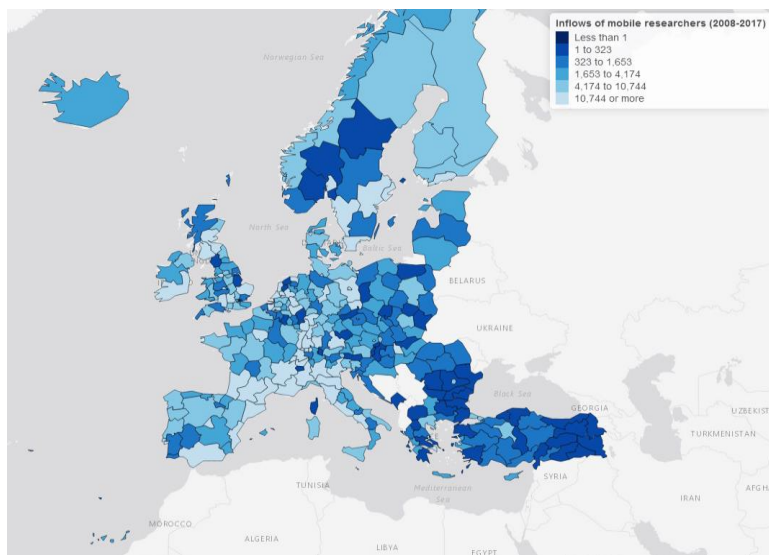

**Figure 1. Total mobility inflows of scientists are shown for NUTS2 regions. In the map, lighter colours represent larger inflows.**

*To what extent are international and regional mobility related?*
Table 1 shows the shares of internationally mobile and regionally mobile researchers in Europe's top ten countries ranked by the share of mobile researchers in each country. As can be seen, the shares of international mobility and regional mobility are significantly different. All countries have higher shares of internationally mobile researchers than regionally mobile researchers.

**Table 1. Share of internationally and regionally mobile researchers for the top 10 countries with the highest share of mobile researchers (2008-2017).**

| Country | % mobility | % international mobility | % regional mobility | # regions |
|---|---|---|---|---|
| 1. Switzerland | 31% | 25% | 6% | 7 |
| 2. Netherlands | 28% | 19% | 10% | 12 |
| 3. Belgium | 27% | 19% | 8% | 11 |
| 4. Germany | 26% | 15% | 11% | 38 |
| 5. Italy | 26% | 16% | 9% | 21 |
| 6. Sweden | 25% | 20% | 5% | 8 |
| 7. Austria | 25% | 21% | 4% | 9 |
| 8. United Kingdom | 24% | 15% | 10% | 40 |
| 9. France | 23% | 14% | 8% | 27 |
| 10. Spain | 23% | 15% | 7% | 18 |

The top countries in the table are also high-income countries separated by relatively short distances, making it easier for researchers to internationalize in those countries. Switzerland has the highest share of mobile researchers (31%), with 25% of mobile scholars having affiliations in at least two countries (internationally mobile) and 6% in at least two regions of Switzerland (regionally mobile). France (23%) and Spain (23%) have the lowest shares of mobile researchers. Switzerland has the highest share of internationally mobile scholars (25%) compared to other countries, whereas Germany (15%), the United Kingdom (15%), Spain (15%), and France (14%) have the lowest share. This can be explained in part by the number and size of these countries' regions (See Figure 2, right panel). France, for example, has more regions (n=27) than Switzerland (n=7) that are larger in spatial scale and scientific workforce. Such conditions may offer researchers a more diverse set of options and point to the role of geography in facilitating and/or constraining scientific mobility.

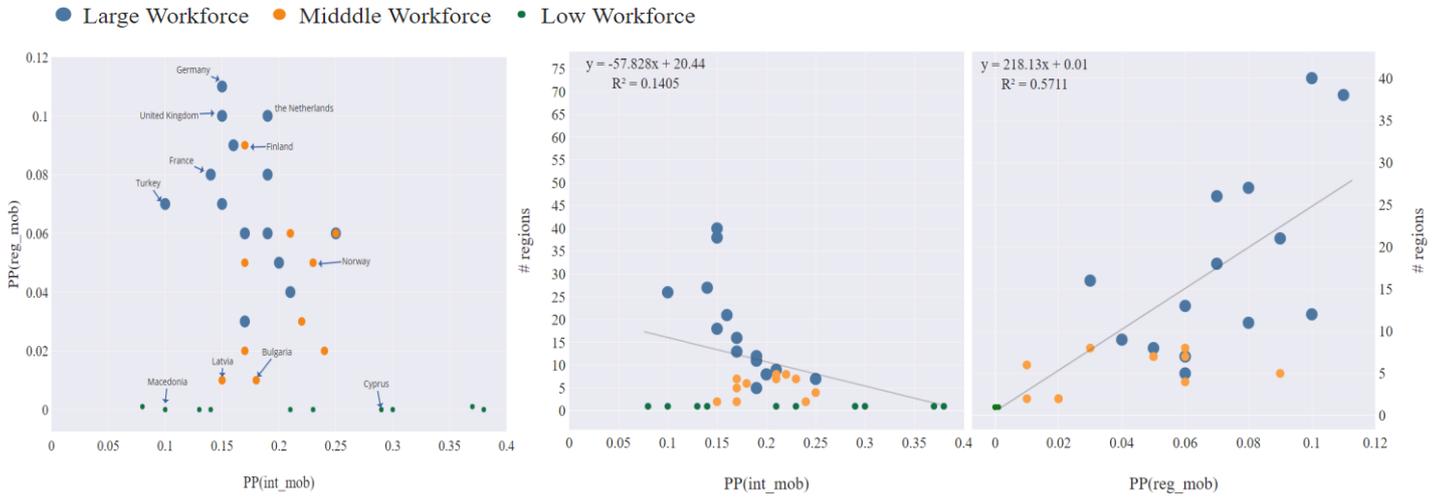

**Figure 2.** Scatterplots of proportion of regional mobility and international mobility by country (2008-2017). Categories are based on country's workforce size (i.e. overall scholars identified in WoS).

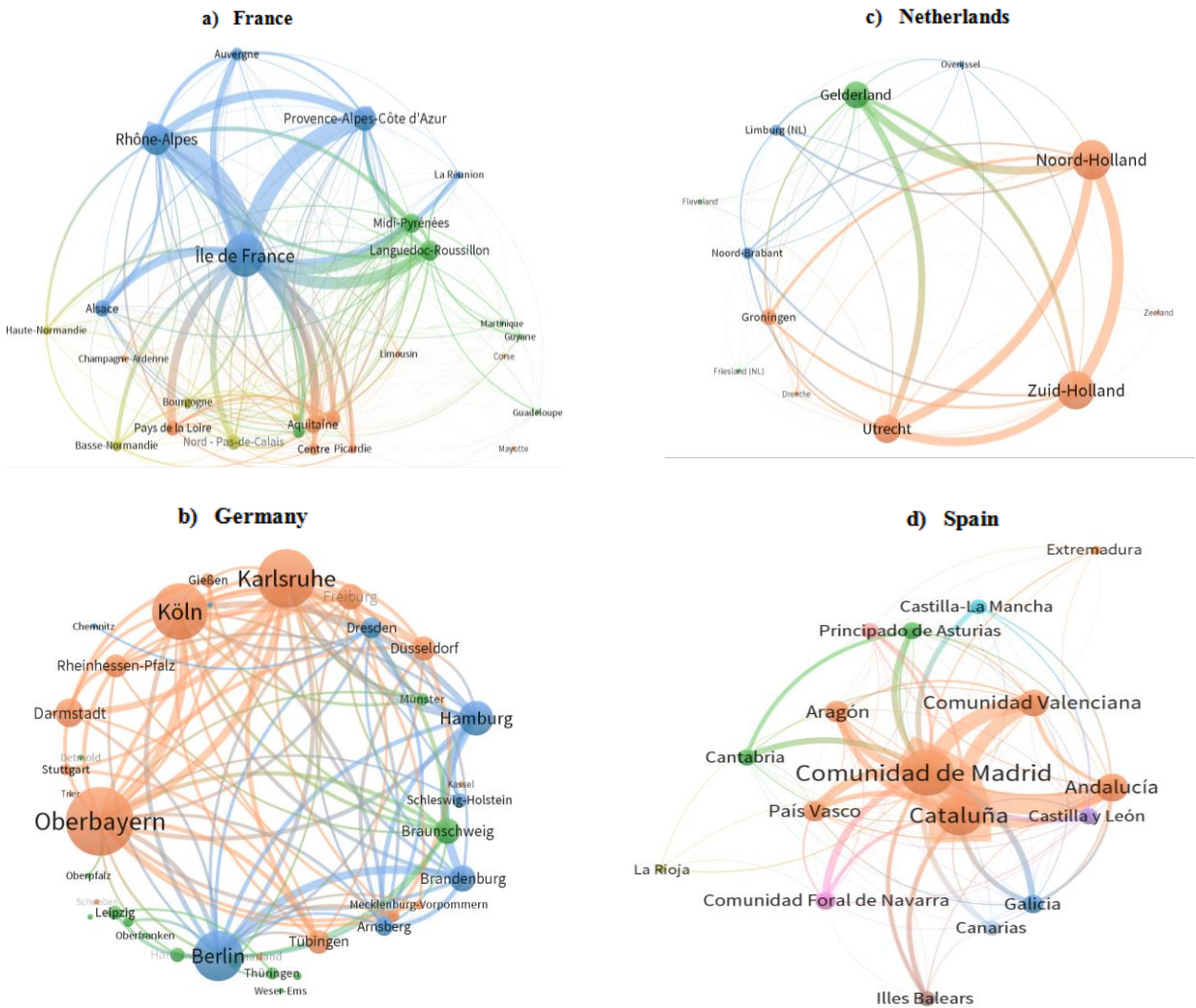

**Figure 3.** Undirected network of regional mobility flows for Germany, Spain, France and the Netherlands (2008-2017). The network visualizations were created using the VOSviewer software (van Eck & Waltman, 2010).

The left panel of Figure 2 suggests that there is no clear relationship between the proportion of scholars who are regionally mobile and scholars who are internationally mobile. The size of countries' workforce (as measured by the total number of scholars publishing in the WoS and identified in the countries) also does not seem to be related to the prevalence of any kind of mobility. As expected, smaller countries with few or no regions (e.g. Cyprus) also do not have any regional mobility. In fact, the two panels on the right side indicate this obvious but noteworthy pattern: larger countries with relatively high number of regions tend to have higher proportions of regional mobility (right panel). The middle panel of Figure 2 shows that larger countries with multiple regions tend to have slightly lower shares of international mobile scholars.

*Which regional mobility network patterns can be seen across countries?*
Figure 3 shows the regional mobility networks of four European countries: France, Netherlands, Germany, and Spain. The nodes represent the different regions in the countries, the size of the nodes is determined by the total number of mobile scholars affiliated with the region, and edges are established by the number of common scholars that have been affiliated to each pair of regions. Although a thorough network analysis is not performed here, the graphs point to two main patterns. Firstly, a concentrated pattern (e.g., France and Spain), in which a few regions (e.g., Ile de France in France and Madrid and Cataluña in Spain) conform strong nodes, having multiple mobility linkages with the other regions in the country. Secondly, a more distributed pattern (e.g. Germany and the Netherlands), in which there are more diverse connected regional nodes (e.g. Oberbayern, Berlin, Köln, Karlsruhe and Hamburg in Germany; and Noord-Holland, Zuid-Holland, Utrecht and Gelderland in the Netherlands), without clearly dominant nodes. Some proximity patterns can also be seen in some of the networks such as the stronger linkages among the northern Spain regions (e.g., Cantabria and Asturias) or among regions in the Randstad area of the Netherlands (e.g., Zuid-Holland, North-Holland and Utrecht). The fact that different networks show very different mobility patterns suggests that national scientific systems in Europe can be organized differently.

**Conclusion and future research**
In this paper, we have presented a first attempt to study mobility patterns among European regions using bibliometric methods. As a proof of concept, it can be concluded that regional mobility is also traceable by bibliometric means. Our preliminary results suggest that mobility patterns between regions differ from mobility patterns between countries (Robinson-Garcia et al., 2019), since they have different incidence across countries. Results indicate that, there is an unequal distribution of regionally mobile scholars across European countries. Moreover, national regional mobility does not seem to have a strong relationship with the level of international mobility of the country. Another conclusion is that, there are two important factors in the consideration of regional mobility: the existence of a regional geographical structure in the country (otherwise regional mobility is not possible), and the existence of a sizeable workforce in the country.

Overall, our results point that regional mobility is a social phenomenon that deserves attention by itself. Thus, these initial observations motivate us to examine the consequences of mobility for the distribution of scientific portfolios (i.e., thematic delineation of regions). As the inflow of researchers into regions is also likely to generate a positive effect on the scientific innovation in receiving locations, we will test whether knowledge generated by scholars in their origin location affects the scientific portfolios of their new location over the years.

We ask, therefore, two interrelated questions: What is the effect of the influx of mobile scientists on the emergence of scientific fields of receiving regions? How stable is the network of inter-regional mobility flows? We will use this approach to identify co-location effects driven by inter-regional publishing. Our expectation is that the international and regional movement of scholars can explain differences in scientific portfolios in European regions. This will include studying the inflow and outflow dynamics of regions in attracting (or sending) scholars; as well as controlling for other demographic aspects such as age and gender of these regionally mobile scholars.

Finally, similar to the identification of internationally mobile researchers by Robinson-Garcia et al. (2019), this study is dependent on the number of publications to identify mobility patterns, the coverage of the database used to extract the publication data, and the completeness of the author-affiliation information. To minimize the influence of these biases, we are currently expanding the methodology to include other document types such as conference publications, book reviews, and letters, and comparing mobility statistics derived from publication data from the Web of Science with highly skilled migration statistics.

**Acknowledgements**
We thank Alfredo Yegros and Wout Lamers for their help in assigning affiliation addresses to the NUTS2 regional classification. This research is partially funded by the South African DST-NRF Centre of Excellence in Scientometrics and Science, Technology and Innovation Policy (SciSTIP) and by the European Commission DG EAC, U-Multirank project.